\documentstyle[12pt,aaspp4,tighten]{article}

\newcommand{\apl}{\lesssim}
\newcommand{\apg }{\gtrsim}
\newcommand{\cmjj}{\mbox{${\rm cm^{-2}}$}}
\newcommand{\hI}{\mbox{${\rm H~I}$}}
\newcommand{\lya}{Ly $\alpha$}
\def\etal{{\it et al.\ }}
\def\ifm#1{\relax\ifmmode#1\else$\mathsurround=0pt #1$\fi}
\def\kms{~{\rm km\ s\ifm{^{-1}}}}

\slugcomment{Submitted to Astrophysical Journal Letters}

\lefthead{Fern\'andez-Soto \etal} 
\righthead{Strong clustering of \lya\ absorbers}

\begin{document}

\title{Strong Clustering of High-Redshift Lyman-alpha Forest Absorption
Systems}

\author{A. Fern\'andez-Soto\altaffilmark{1,2}, K. M. Lanzetta\altaffilmark{3},
X. Barcons\altaffilmark{1}, R. F. Carswell\altaffilmark{4},J. K.
Webb\altaffilmark{5}, A. Yahil\altaffilmark{3}}

\altaffiltext{1}{Instituto de F{\'\i}sica de Cantabria, Consejo Superior de
Investigaciones Cient{\'\i}ficas -- Universidad de Cantabria, Facultad de
Ciencias, 39005 Santander, SPAIN.}

\altaffiltext{2}{Departamento de F{\'\i}sica Moderna, Universidad de Cantabria,
39005 Santander, SPAIN.}

\altaffiltext{3}{Astronomy Program, Department of Earth and Space Sciences,
State University of New York at Stony Brook, Stony Brook, NY 11794--2100,
U.S.A.}

\altaffiltext{4}{Institute of Astronomy, University of Cambridge, Cambridge CB3
0HA, U.K.}

\altaffiltext{5}{School of Physics, University of New South Wales, P.O. Box 1,
Kensington, NSW, AUSTRALIA}

\begin{abstract} 

  We use new observations of very weak CIV absorption lines associated with
high-redshift \lya\ absorption systems to measure the high-redshift \lya\ line
two-point correlation function (TPCF).  These very weak CIV absorption lines
trace small-scale velocity structure that cannot be resolved by \lya\
absorption lines.  We find that (1)~high-redshift \lya\ absorption systems with
$N(\hI) > 3 \times 10^{14}$ \cmjj\ are strongly clustered in redshift,
(2)~previous measurements of the \lya\ line TPCF underestimated the actual
clustering of the absorbers due to unresolved blending of overlapping velocity
components, (3)~the present observations are consistent with the hypothesis
that clustering of \lya\ absorption systems extends to lower column densities,
but maybe with smaller amplitude in the correlation function, and (4)~the
observed clustering is broadly compatible with that expected for galaxies at $z
\sim 2-3$.  We interpret these results as suggesting that many or most \lya\
absorbers may arise in galaxies even at high redshifts, and, therefore, that
the \lya\ forest probes processes of galaxy formation and evolution for
redshifts $z \lesssim 5$.

\end{abstract}


\section{Introduction}

  The observational result that high-redshift \lya\ absorption systems appear
not to cluster strongly in redshift (e.g.  Sargent \etal  1980) has driven
most discussion about the origin of the \lya\ forest.  This result has
generally been interpreted as evidence that high-redshift \lya\ absorbers arise
in intergalactic clouds rather than in galaxies.  Recent studies of the
relationship between \lya\ absorbers and galaxies at redshifts $z \apl 1$,
however, directly demonstrate that many or most low-redshift \lya\ absorbers
(or at least those satisfying $W_{rest}({\rm Ly}\alpha ) \apg 0.3$\AA) arise in
galaxies rather than in intergalactic clouds (Lanzetta \etal  1995).  Why is
it that \lya\ absorbers appear not to cluster strongly in redshift whereas
low-redshift \lya\ absorbers appear to arise in galaxies?

  One suggestion is that there exist two distinct populations of \lya\
absorbers: a rapidly evolving, unclustered, intergalactic population that
dominates at high redshifts, and a slowly evolving, clustered, galactic
population that dominates at low redshifts (e.g.  Bahcall \etal 1995).  Another
possibility is that previous measurements of the high-redshift \lya\ two-point
correlation function (TPCF) have underestimated the actual clustering of the
absorbers---presumably due to unresolved blending of overlapping velocity
components---and \lya\ absorbers arise in galaxies at all epochs.

  Here we examine the second of these possibilities, that previous measurements
of the high-redshift \lya\ TPCF have underestimated the actual clustering of
the absorbers, using new observations of very weak CIV absorption lines
associated with high-redshift \lya\ absorbers (Cowie \etal 1995, hereafter
CSKH), \S2.  These very weak CIV absorption lines trace small-scale velocity
structure that cannot be resolved by \lya\ absorption lines because (1)~the
atomic weight of C is 12 times the one of H, so the thermal broadening of CIV
absorption lines is 3.5 times smaller than that of \lya\ lines, and (2)~CIV
absorption lines suffer far less saturation because of the difference in column
densities.  We show that the CIV lines indeed help to reveal the underlying
velocity correlation of the \lya\ systems, and that this same velocity
structure is blended away in the \lya\ data, \S3.  We conclude with a
comparison of the derived velocity clustering of the \lya\ absorbers with that
of galaxies at the present epoch, \S4.

\section{Data}

  The observations by CSKH consist of high spectral resolution (${\rm FWHM}
\approx 8$ \kms), high signal-to-noise ratio ($S/N \approx 50$ per resolution
element) spectra of three QSOs obtained with the Keck telescope and the HIRES
spectrograph.  The observations generally cover both the \lya\ and the
corresponding CIV wavelength regions and are sensitive to CIV absorption lines
arising in CIV column densities as low as $N({\rm CIV}) \approx 10^{12}$ \cmjj.

  From the observations, CSKH selected a complete sample of 38 \lya\ absorption
lines satisfying $N(\hI) \ge 3 \times 10^{14}$ \cmjj.  They then eliminated
seven of these absorption lines due to contamination by unrelated metal
absorption lines or lack of coverage of the corresponding CIV wavelength region
or because the lines produce corresponding Lyman-limit absorption (which
indicates $N(\hI) \apg 2 \times 10^{17}$ \cmjj).  The resulting sample thus
contains 31 \lya\ absorption lines satisfying $3 \times 10^{14} \ \cmjj \le
N(\hI) \le 2 \times 10^{17} \ \cmjj$.  For each member of this sample, they
searched the corresponding CIV wavelength region for CIV absorption lines and
applied a Voigt profile fitting procedure to the identified CIV absorption
lines to measure redshifts, Doppler parameters, and column densities.

  Here we use the absorption system parameters derived by CSKH in their profile
analysis, which are summarized in their Table 1a.  The average redshift of the
absorbers is $\langle z \rangle = 2.6$, the median column density of the
absorbers is $N(\hI) = 8.1 \times 10^{14}$ \cmjj, and the typical CIV/HI ratio
of the absorbers is $3 \times 10^{-3}$.  Of the final sample of 31 \lya\
absorption lines, 15 are observed to have associated CIV absorption, of which
six show small-scale velocity structure with between two and nine velocity
components per \lya\ absorption line.

\section{Analysis}

\subsection{High-Redshift \lya\ Two-point Correlation Function}

  Our primary assumption is that very weak CIV absorption lines trace
small-scale velocity structure that cannot be resolved by \lya\ absorption
lines.  Hence the goal of the analysis is to measure the high-redshift \lya\
TPCF by using very weak CIV absorption lines instead of the \lya\ absorption
lines themselves.

  To do this we use the results summarized in Table 1a of CSKH.  In cases where
CSKH identified one or more CIV absorption lines with a single \lya\ absorption
line, we use the redshifts of all CIV absorption lines in the analysis.  In
cases where CSKH identified no CIV absorption lines with a single \lya\
absorption line, we use the single redshift of the \lya\ absorption line in the
analysis.  This procedure yields a total of 52 absorption redshifts.  We then
use these absorption redshifts to construct the \lya\ line TPCF by normalizing
the distribution of velocity pairs with respect to an unclustered distribution
of redshifts.

  The results are shown in Figure 1, which plots in the upper panel the
high-redshift \lya\ line TPCF as traced by CIV absorption lines.  (The error
bars shown in Figure 1 are based on a modified ``bootstrap'' technique that
yields approximately correct results even for correlated data.  Details of this
technique will be presented elsewhere.)  It is clear from Figure 1 that the
high-redshift \lya\ TPCF indicates very strong clustering on velocity scales
$\lesssim 250$\kms.  We therefore conclude that high-redshift \lya\ absorption
systems with $N(\hI) > 3 \times 10^{14}$ \cmjj\ are strongly clustered in
redshift.

\subsection{Blending of Overlapping Velocity Components}

  The results of \S3.1 demonstrate that high-redshift \lya\ absorption systems
with $N(\hI) > 3 \times 10^{14}$ \cmjj\ are strongly clustered in redshift,
whereas all previous analyses have found that they are
either unclustered (Sargent \etal  1980) or only very weakly clustered in
redshift (e.g.  Webb 1987; Barcons \& Webb 1991).  How are these results
compatible?

  To examine this issue, we apply the standard method of measuring the \lya\
TPCF to models of the \lya\ absorption lines observed by CSKH.  We first
generate a set of \lya\ absorption lines according to the results
in Table 1a of CSKH.  We adopt a constant CIV/HI ratio of $3 \times
10^{-3}$ and assume that the Doppler parameters are due to thermal motions,
convolve the synthetic absorption lines with the appropriate instrumental
response and add noise to match the actual signal-to-noise ratio of the
observations.  Next, we fit the resulting synthetic spectra using the Voigt
profile fitting routine described previously by Lanzetta \& Bowen (1992).  For
each absorption line we add velocity components until the decrease in $\chi^2$
is smaller than the accompanying decrease in degrees of freedom, $\nu$.
Finally, we construct the \lya\ TPCF according to the procedures described in
the previous section, but this time using the fitted redshifts instead of the
actual redshifts.

  The results are shown in Figure 1, which plots in the lower panel the
high-redshift \lya\ TPCF as traced by \lya\ absorption lines.  It is clear that
the \lya\ absorption lines cannot reveal the strong clustering indicated by the
CIV absorption lines.  The lower panel of Figure 1 may be directly compared
with the high-redshift \lya\ TPCF presented by Hu \etal (1995); both use
observations of nearly the same quality, and both obtain practically identical
results.  Note that our stopping criterion for velocity components,
$\Delta\chi^2<\Delta\nu$ purposely allows even marginally significant lines to
be included.  If no correlation is obtained even with this generous criterion,
it certainly will not be found with a more conservative one.  We therefore
conclude that previous measurements of the high-redshift \lya\ TPCF have
underestimated the actual clustering of the absorbers due to unresolved
blending of overlapping velocity components.

\begin{figure}
\plotone{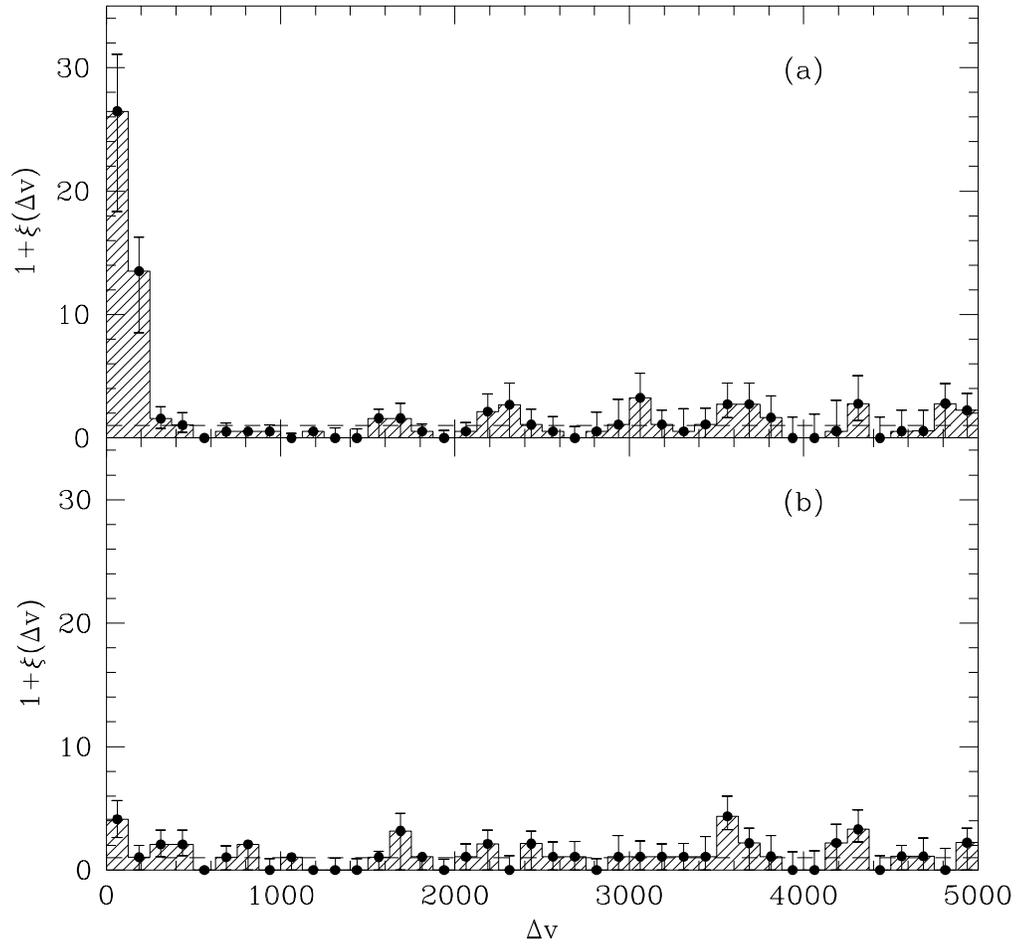}
\caption{High-redshift \lya\ TPCF as 
traced by very weak CIV absorption lines (upper panel) and as traced by \lya\ 
absorption lines (lower panel). \label{fig1}}
\end{figure}

\subsection{Extension to Lower Column Densities}

  The results of the previous section demonstrate that previous measurements of
the high-redshift \lya\ TPCF of absorbers with $N(\hI) > 3 \times 10^{14}$
\cmjj\ have underestimated the actual clustering of the absorbers.  Can this
result extend to lower column densities for which blending is presumably
weaker?

  To examine this issue, we repeat the analysis described in the
previous section for \lya\ lines generated in two different ways.  For
the first simulation we assume that the CIV Doppler parameter $b$ is
entirely due to thermal motion, so $b(HI)=\sqrt{12}b(CIV)$, and reduce
the HI column densities by a factor of 100 with respect to the
original ones.  An example of just how a Voigt profile fit to high
spectral resolution, high signal-to-noise ratio observations can
underestimate the actual number of velocity components comprising an
absorption line is shown in Figure 2, in which panel (a) shows the
result of synthesizing the complex of lines at $z=2.7853$ toward
Q0302$-$003 (with HI column densities decreased by a factor of 100
with respect to the original ones), panel (b) shows the
actual components of the \lya\ absorption line, and panel
(c) shows the result of the Voigt profile fitting procedure.  This
complex of nine lines is adequately fitted ($\chi^2 / \nu = 0.91$)
with only three velocity components.  The derived spectrum is then
fitted in the same way used in \S3.2.  The resulting TPCF, Figure 3a,
is still weaker than that of the CIV lines, but clearly detectable.

\begin{figure}
\plotone{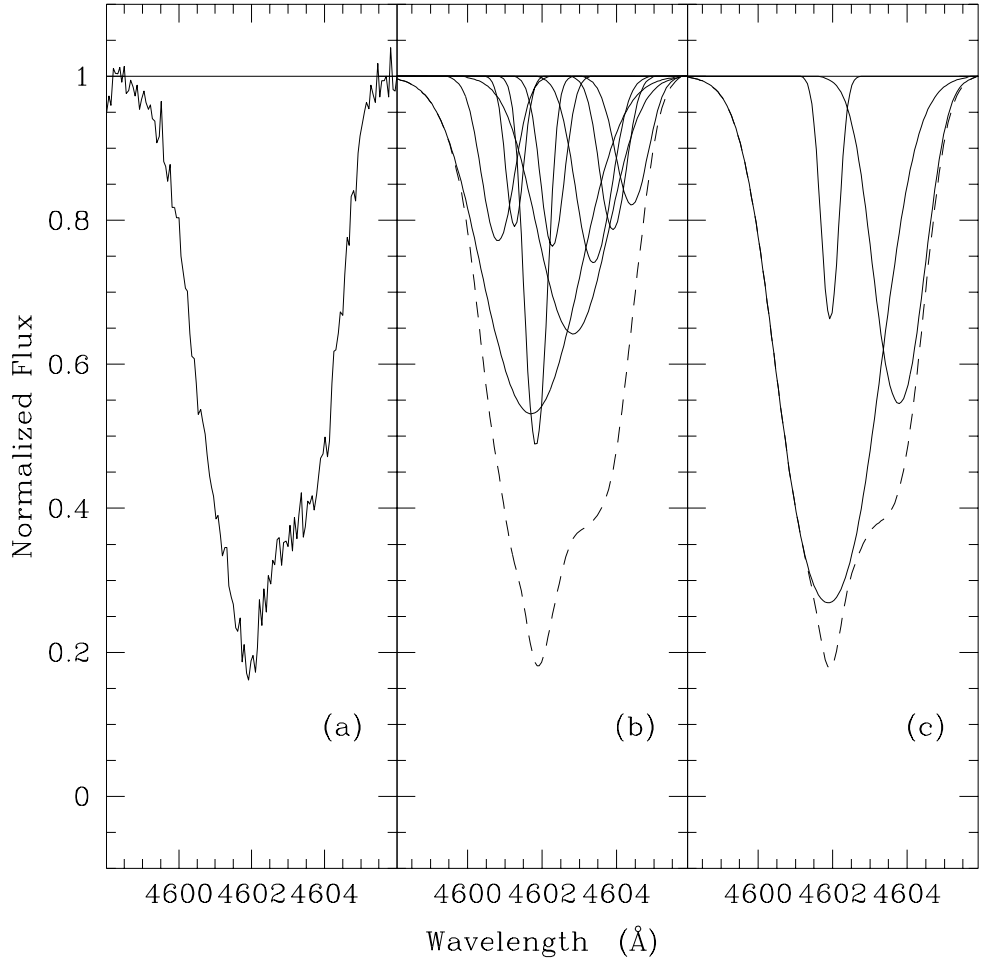}
\caption{Example of how a Voigt profile fit to high spectral 
resolution, high signal-to-noise ratio observations can underestimate the 
actual number of velocity components comprising an absorption line.  Panel (a) 
shows the result of synthesizing the complex of lines at $z=2.7853$ toward 
Q0302$-$003 (with HI column densities decreased by a factor of 100 with 
respect to the original column densities), panel (b) shows the actual 
components that comprise the \lya\ absorption line, and panel (c) shows the 
result of the Voigt profile fitting procedure. \label{fig2}}
\end{figure}

  The assumption that all the velocity dispersion is thermal leads to
temperatures in excess of $6\times 10^4$K in some cases, and this is
inappropriate in most models (see Charlton, 1995, for a review of the models).
We therefore add a second simulation in which the temperature is assumed to be
$2\times 10^4$K, and any excess Doppler parameter is ascribed to turbulence and
applied equally to the CIV and \lya\ lines.  In a few cases the CIV Doppler
parameter is just below the assumed thermal value, and in these cases we simply
adopt the $2\times 10^4$ thermal width for {\lya}.  In this simulations the HI
column density is assumed to be $10\times$ that for CIV for each component.
The \lya\ lines are now generally narrower than in the
first simulation, and so the component structure is more easily detected.
Consequently, the \lya\ line TPCF will have larger values at low velocity
separations, as can be seen from Figure 3 (b).  This should be compared with
the observational result that little clustering is found at these redshifts
(e.g.  Rauch \etal, 1992).

  These simulations are indeed too na\"{\i}ve, as we are not taking into
account the increment in line number density at low column densities.  This
increment will produce strong blending effects among low column density lines
themselves and also with the higher column density lines.  Their combined
effect is very difficult to simulate, as it depends very strongly on the
higher-order correlation functions of the distribution of the lines.  In
addition, there are observations that suggest that the amplitude of the
clustering is smaller at these low column densities (Hu \etal 1995).  All of
these effects could very well erase all the signal in the correlation function
for low column density lines, and hence we cannot conclude anything on the
behavior of these low column density lines other than it is compatible with
being clustered but maybe with a smaller clustering amplitude.

\begin{figure}
\plotone{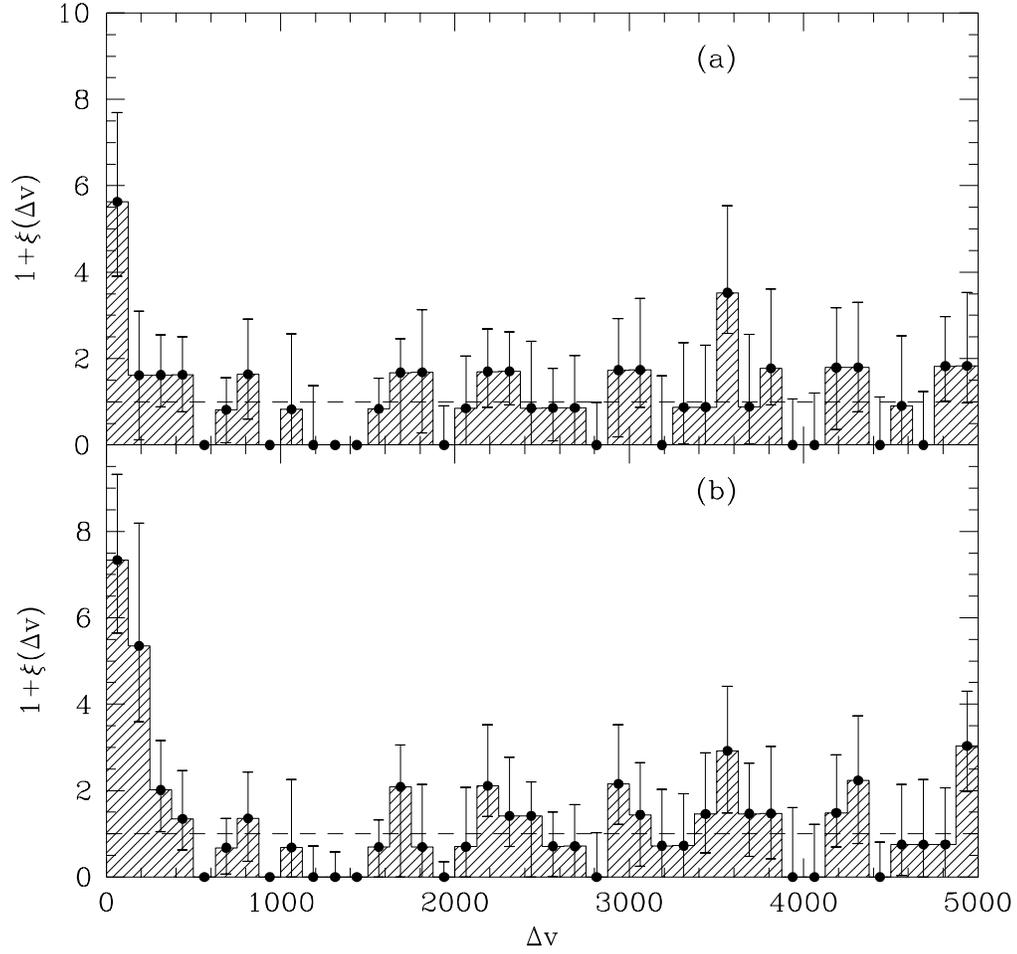}
\caption{The TPCF for low column 
density clouds as obtained with two different models: Thermal-only broadening 
(panel a) and thermal plus turbulence (panel b). \label{fig3}}
\end{figure}

\section{Discussion and Summary}

  The most significant result of the previous sections is that high-redshift
\lya\ absorbers with $N(\hI) > 3 \times 10^{14}$ \cmjj\ are strongly clustered
in redshift on velocity scales $\apl 250 \kms$.  While the effect might be due
to pairs of clouds with small velocity differences causing the observed TPCF
(Miralda-Escud\'e \etal 1995; Rauch 1995), we could be seeing real clustering.
With the observed velocity correlation length we can not decide whether the
\lya\ absorbers are independent entities, as has generally been assumed so far,
or clouds within the halos of galaxies, the possibility we are exploring here.
More detailed questions are even harder to answer, for example the type of
galaxies in which the absorbers might reside, whether we are observing multiple
clouds within the same galaxies, and the ionization state of carbon in the
clouds.  We can only ask if the strength of \lya\ clustering is consistent with
expectations of galaxy clustering at these early epochs.

  To examine this issue, we consider a simple model for the evolution of the
galaxy TPCF.  In a first step, we ignore peculiar motions and motions of clouds
within galaxies and assume that as a function of velocity and redshift the
galaxy TPCF can be described by (Efstathiou \etal 1991)
\begin{equation}
\xi(v,z) = (1 + z)^{-3-\epsilon} \left[ \frac{v}{r_0 H(z)} \right]^{-1.8}
\end{equation}
where $H(z)$ is the Hubble constant at epoch $z$ (we take $q_0=0.5$) and
$H_0r_0=550\kms $ is the present-day galaxy correlation length.  The
evolutionary parameter $\epsilon$ takes the value $-1.2$ for comoving
structures, 0 for virialized clusters, and 0.8 for linearly growing
perturbations.  Recent theoretical studies (Hamilton \etal  1991; Jain \etal
1995) show a steeper dependence on redshift at intermediate stages between the
linear and virialized limits.

  To avoid the divergence of this function at small values of $v$, we take
$\xi(v,z)$ to be constant below a given velocity difference $v_0$ and equal to
$\xi(v_0,z)$.  We also convolve it with a Gaussian distribution with width
$\sigma$ to account for random motions.  In this way we get a set of different
models defined by three parameters, $\epsilon$, $\sigma$, and $v_0$, which we
allow to vary within the limits: $-2 < \epsilon < 4$, $0 < \sigma < 500 \kms$,
and $1 < v_0 < 80 \kms$.

   Predictions of all these models are then compared with the observed TPCF of
high-redshift \lya\ absorbers.  The best fit is achieved for
$\epsilon= 2.4$ and $\sigma=100$ \kms .  The 1-, 2- and 3-$\sigma$ confidence
regions obtained using $v_0$ as an uninteresting parameter are plotted in
Figure 4.  From this calculation it is clear that if normal galaxies host the
\lya\ absorbers, at an average rate of one per galaxy, their correlation
function is evolving rapidly and the combined intragalactic and intergalactic
velocity dispersion is $\lesssim 150\kms$.

\begin{figure}
%\plotone{fsclust4.eps}
\vbox to14cm{\rule{0pt}{14cm}}
\includegraphics{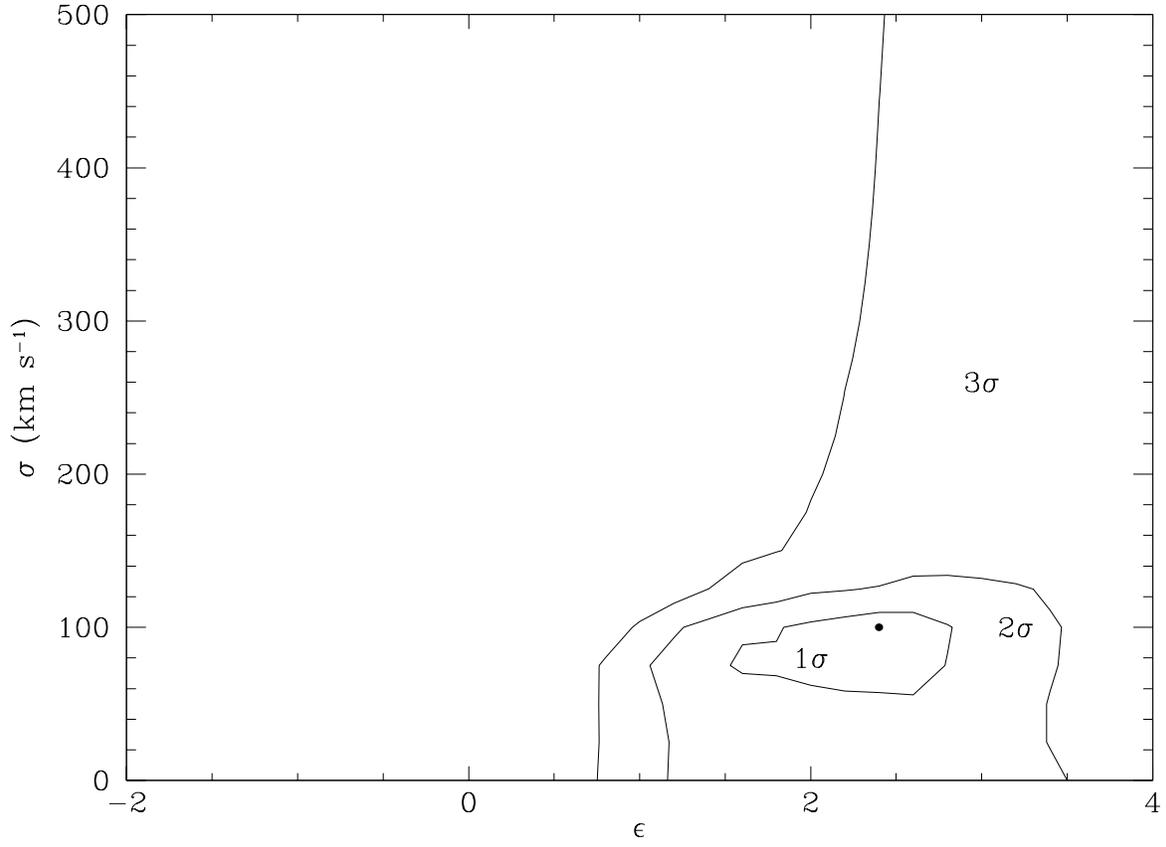}
\caption{Confidence limits in the parameter space formed by 
the clustering evolution parameter $\epsilon$ and the typical velocity of 
galactic halo motions $\sigma$. $1-, 2-$ and $3\sigma$ confidence contours 
are plotted. \label{fig4}}
\end{figure}

  Note that our redshift range is relatively small, $\Delta z \approx 0.6$, so
that we can not separately fit $r_0$ and $\epsilon$.  Our determination of
$\epsilon$ is therefore anchored by the general galaxy correlation function at
the present epoch.  This may be inappropriate in several respects.  We
overestimate the correlation function if the host galaxies of the \lya\ clouds
are less clustered at the present epoch, for example if they are mostly spiral
galaxies, or we underestimate it if there are multiple \lya\ absorbers in
galaxies at high redshift.  The best we can deduce from our simple analysis is
that the observed clustering of high-redshift \lya\ absorbers is broadly
consistent with the expected clustering of galaxies.

We conclude that (1)~High-redshift \lya\ absorbers with $N(\hI) > 3 \times
10^{14}$ \cmjj\ are strongly clustered in redshift on velocity scales $\apl
250$ \kms, (2)~Previous measurements of the \lya\ TPCF have underestimated the
actual clustering of the absorbers due to unresolved blending of overlapping
velocity components, (3)~The present observations may be consistent with the
hypothesis that clustering of \lya\ absorption systems persists to lower column
densities, being likely that the clustering is smaller at low column densities,
and (4)~The observed TPCF is broadly compatible with that expected from
galaxies at $z \sim 2-3$.

  We interpret these results to suggest that many or most \lya\
absorbers may arise in galaxies at all epochs, and therefore that the
\lya\ forest probes the processes of galaxy formation and evolution
for redshifts $z \lesssim 5$.

\acknowledgments

  AFS thanks E.  Mart\'{\i}nez-Gonz\'alez for useful discussions and an
anonymous referee for calling our attention on the gravitational collapse
model.  AFS acknowledges support provided by a Spanish MEC studentship.
Partial financial support to AFS and XB is provided by the Spanish DGICYT under
project PB92--0741, KML is supported by NASA grant NAGW--4433 and by a Career
Development Award from the Dudley Observatory, and AY by NASA grant NAG--51228.

\end{document}